# Univariate and Multivariate LSTM Model for Short-Term Stock Market Prediction

Vishal Kuber, Divakar Yadav, Arun Kr Yadav

20mcs118@nith.ac.in, divakaryadav@nith.ac.in, ayadav@nith.ac.in

Department of Computer Science & Engineering, NIT Hamirpur (HP), India

**Abstract-** Designing robust and accurate prediction models has been a viable research area since a long time. While proponents of a well-functioning market predictors believe that it is difficult to accurately predict market prices but many scholars disagree. Robust and accurate prediction systems will not only be helpful to the businesses but also to the individuals in making their financial investments. This paper presents an LSTM model with two different input approaches for predicting the short-term stock prices of two Indian companies, Reliance Industries and Infosys Ltd. Ten years of historic data (2012-2021) is taken from the yahoo finance website to carry out analysis of proposed approaches. In the first approach, closing prices of two selected companies are directly applied on univariate LSTM model. For the approach second, technical indicators values are calculated from the closing prices and then collectively applied on Multivariate LSTM model. Short term market behaviour for upcoming days is evaluated. Experimental outcomes revel that approach one is useful to determine the future trend but multivariate LSTM model with technical indicators found to be useful in accurately predicting the future price behaviours.

## 1. Introduction

Financial stock market forecasting is among the most critical problems in computer science today. It results from factors such as physical traits vs the mind, rational behaviour vs logic, etc. These factors combine to produce stable values that are extremely difficult to predict accurately.

Long-term research examines the financial timeline and forecasts future prices and market movement. Although some research indicates a hypothetical market hypothesis and believes that quotations are incorrect, there are suggestions in the literature that careful coding can be used to predict stock prices with the highest level of accuracy. Models are hypothesised. The guessing model's accuracy was also dependent on the set of variables used to model the model, the algorithms used, and how the model was developed. The documents contain recommendations for the eventual collapse of the stock price chain. Machine learning programmes and in-depth learning methods are also popular in stock price analysis and forecasting. Artificial intelligence is a branch of computer science and technology that has advanced significantly in the last few years. It has simplified a significant number of real-world issues. It tries to mimic human behaviour, such as how people learn new things over time and use their inventiveness to improve. Nobody imagined a small driver's car, tangible assistance, and the list goes on. One of the problems that AI still needs to solve with great accuracy is 'Asset Prediction.' Many people from the industry have been drawn to higher education by stock forecasting. Stock forecasting is not like traditional prediction problems. Forecasting the weather or predicting a specific attribute, for example, is far easier than forecasting stock. Many researchers believe that predicting stock prices is akin to giving everyone a printing press. Anyone or any organisation can use it to predict prices and assist them in making money and surviving market risks. It is not the same as doing nothing about it. Many researchers put forward algorithms, which are suggested structures for improving predictive accuracy. Several machine learning algorithms have been used to solve this problem, including regression, easy measurement, moving averages, genetic algorithms, and SVM, but none provide significant accuracy for each. As a result, different problem modes with the same mixed methods are provided. Approaches for solving problems have been around for a quite long time. They are considered among the most challenging problems to address in the data science business. This involves a significant number of issues. One needs to forecast the prices and identify the patterns in stock market data. We will look at one such method with LSTM or temporary memory. The recurrent neural network (RNN) is a powerful time-series data processing method. LSTMs are inferior to feed transmissions and sensory RNN networks in many ways. This is because they have a commemorative property by selecting long patterns. Some RNN models have been used to forecast stock market movements. One of the most successful RNN structures is LSTM.

*1.1 Why Stock Prediction is Necessary*

The stock market is behaving strangely. As financial markets have frequent rise and falls, the stock values moves independently at all times. Because of random turbulence in the financial market, it becomes difficult to create data history. Furthermore, Mbazi et al. emphasised that a number of variable can have an impact on the market such as business and economic conditions, as well as political and personal issues. Because the stock market is so volatile, it is impossible to forecast stock price fluctuations.

Because of globalisation and the advent of information and communication technology, many people flock to the stock market in search of high returns in an easy-to-invest-in area (ICT). As a result, stock market forecasts have emerged as a critical issue for investors. The two types of stock market prediction approaches are fundamental approach and technical approach. Fundamental approach is the process in which analysing all aspects of a company's internal value, whereas in technical approach we predict the future price values through graph analysis. When basic analysis is used, other issues may arise. For example, time forecasts may be reduced, compliance may be jeopardised. Due to the limitations of basic analysis, many studies in stock market forecasting rely on technical analysis.

*1.2 Importance of Time Frame*

Time is a useful tool for examining a pattern over time. This data is used by investors to forecast future market movements. There are numerous frameworks available, with daily being the most popular. In this essay, we'll discuss why it's critical to invest in this technique. Regardless of whether a fixed weekly, monthly, or hourly schedule is available, most retailers appear to operate on a daily basis.
   a) Intraday trader (day trader) - A 60-minute period is used to learn about the main trend, followed by 5- or 15-minutes set aside for planning a short-term process. The sellers here buy and sell in a single day.
   b) Swing Trader- A trader in this category organizes his positions on a daily basis. The weekly chart can be used to plan the major trends as well as the time limit of the 60-minute short-term trend. The swing dealer usually takes his position and squares during the week.
   c) Long-term Trader- Long-term traders are interested in trading for the long term. They primarily rely on weekly and monthly meetings to maintain their positions. As a result, before you embark on your next adventure, think about the significance of each time period.

*1.3 Back testing of Stocks*

A background check, based on previous data, determines how a trading strategy or price model will perform in the real world. The basic premise is that any method that has previously worked will most likely work in the future, and any strategy that has not previously worked will most likely work in the future. It is critical to set the historical data time for testing purposes when checking for historical data views. If it works, it can be tested, or external data can be used to confirm its dynamic performance.

The financial market is an important place not only for companies to raise capital, but also for individuals to make a profitable investment. As a result, stock forecasting has always been a difficult subject to research. Because stock data is complex and versatile, it is critical to have a thorough understanding of all of its components. Otherwise, you risk losing your investment.

The proposed study focuses on short term future price behavior of two selected Indian stocks namely Reliance and Infosys. The key rationale for picking these two companies is because they are both heavily traded and hence will reflect the Indian economy as a whole. In the NIFTY 50 index, Reliance has a weightage of 11.89% and Infosys has a weightage of 9.13%, making them the top two contributors. NIFTY 50 is an Indian stock market index that measures the average of 50 of India's major firms listed on the National Stock Exchange (NSE) (National Stock Exchange). All the data required for the study is taken from yahoo finance site. A total of ten years of data from the past from (2012-2021) is used. The proposed study focuses on two approaches in the first one, closing values are given as an input to LSTM model. This approach will be able to determine upcoming tread i.e. whether the price will go up or go down. In the second approach, technical indicators values are calculated from the closing values and then fed to multivariate LSTM model. This method help us to know about upcoming prices for the selected stock.

The rest of the study is presented with the following sections: section 2 explains literature review, where we summarizes the previous studies carried out with respect to the proposed topic. This section is necessary because it gives us idea about the previous research in the topic until now and what are the drawbacks and weaknesses in previous studies. Section 3 explains about the research data. It gives us idea about the datasets, attributes and explains each term related to it. In the section 4, methodology and implementation, it has explained about the LSTM model and implementation approaches we have used. Section 5 shows the experimental results. In the last, section 6, concludes the proposed study.

## 2. Literature Review

Before moving forward with any additional research, it is critical to first comprehend the basic studies that the researchers have completed and published. Patel et al. [16] use four different speculative model to investigate two strategies for incorporating these species: Vector Support Machine (SVM), Artificial Neural Network (ANN), Random Forest (RF) and Naive-Bayes. The first method of data entry identifies ten technical constraints using stock trading data, whereas the second method concentrates on replicating these parameters in the form of static decision data. The accuracy of each speculation model is tested for each of these two input methods. From 2003 to 2012, the study used 10-year historical data from two stocks, Reliance Industries and Infosys Ltd. The S&P Bombay Stock Exchange (BSE) Sensex and the CNX Nifty are two Indian markets. Depending on the test results, the random field outperforms the other three speculative models in full performance in the first data entry method, ten technical values are defined as continuous values in this model. Depending on the test results, all predicted models perform better when these technological frameworks are represented as data-determining trends. In this paper's Deterministic Trend Data Preparation Structure, the concept of stock price technical indicators is classified as 'upwards' or 'downwards.' Many categories, such as "more chances to climb," "more chances to fall," "less chances to climb," "less chances to fall," and "middle signal." should be considered. 'This may provide a more accurate input into the unpublished professional editing engine, namely the speculative algorithms of this paper,' the researcher explained.

Similarly, the same authors Patel et al. [17] published a study in which hybrid models such as SVR–ANN, SVR–RF, and SVR–SVR were stated. The hybrid approach presented in [17] clearly outperforms the individualistic models presented in [16].

In this study, Zhang et al. [24] concentrated on assessing stock trends and forecasting future stock values changes. Following that, a new method named in the status box is suggested. In contrast to the predictive variable conversion function, Stock units are split into three sorts of boxes using the status box approach, each reflecting a particular detail. A variety of ML methods also helpful to categories such boxes in order to determine if particular box conforms to trend and to predict stock price trends according to box location.

Chen and Hao [4] look at stock market indicators differently, which is an interesting and important lesson in investment and use because the best trading strategies can be most profitable while minimizing risk. Various methods, including mechanical learning programs, have been tried and tested in order to obtain accurate predictions. In this paper, we present a comprehensive integrated vector weight machine with a K-neighbor feature that can predict stock market indicators accurately. Create a detailed theory of the SVM data classification feature, which provides different weights for different parameters based on segmentation values, to get started. Then, in order to gain weight, we subtract information from each element's value. They use the K-neighborhood feature in the archives to predict stock market trends in the future by installing nearby computers. Finally, the test results of the Chinese stock market indicators popularly known as the Shanghai and Shenzhen stock exchanges were presented in order to assess the efficacy of our proven method. In the small, medium and longer term, our recommended model can outperform the 'Shanghai Stock Exchange Composite Index' and the 'Shenzhen Stock Exchange Component Index'. The suggested techniques is also be used to create a different financial market indicator.

Chong et al. [5] provide a comprehensive and objective evaluation of the benefits and drawbacks of a thorough examination of stock market analysis and forecasting methodologies. They compared the results of three uncontrolled techniques for the release of key component analysis features, auto encoder, and Boltzmann machine confined to a general network capability to predict future market performance using high-level 'intraday stock' recovery as input data. Deep neural networks, according to Royal's findings, able to retrieve new information from the leftovers of a self-harming model and improve forecast performance; however, when a stress model is used for network residues, this is not the case. When a forecasting function is used in the analysis of market structure based on covariance, covariance estimates improve significantly. This study provides useful information and guidelines for future research into how extensive research in share market modeling and forecasting, network might be quite useful. They propose an in-depth aspect of the stock market forecast model in this paper. We generate seven sets of symbols by analyzing key components, auto encoders, and Boltzmann's limited range, as well as building three Deep Neural Networks (DNN) to forecast forthcoming investment return, ranging from green standardization to debt recovery. It was used in the 'Korean stock market' and you discovered that DNNs outperform the standard exercise approach in training, but the advantages disappear dramatically in testing.

Nam and Seong [15] present unique research approach in predicting stock values changes depending on capital risk estimates in this article. Our approach, in particular, investigates the causal relationships between companies and takes into account the direct impact of the General Divisions of Industrial Division. To find the cause, the proposed method employs entropy transmission and reads multiple letters to combine the characteristics of the target company and the causal variable. Our test results show that the framework based on causal analysis outperforms the previous two algorithms, based on 'Korean market' data's external testing. Furthermore, the test results indicated that the proposed strategy is effective and can accurately forecast values even if there isn't any finance information is available for the target company but financial news is available for the causative entities. Their discovery implies that identifying causal relationships, as well as developing machine learning algorithms and establishing well-developed hypothetical links such as complicated concept systems, is important in predictive problems. Analyze causal relationships between non-regulated firms by analyzing individual companies and within the sector in this study to improve model accuracy. This method improves prediction by exceeding the critical limit of previous research, which is regarded as a dual target within the GICS areas. We use the entropy transmission method, which is widely used in physics, to clearly analyze the causal relationship and apply it to the predictive model.

Chen and Hao [3] presented research on trademark prediction, an interesting but difficult business topic in research space because share market is a flexible as well as critical system with many complex connections, and even minor improvements in forecasting performance can be profitable. A variety of methods, including innovative methods to attract the attention of investors and analysts, have been developed to ensure the availability of trademarks. In this study, they used a vector measurement device (WSVM) to introduce a complete and effective method of integrating key component processing (PCA) into the weight trading platform (PCA-WSVM). First, as a problem, the stock forecasting model is divided into four categories. In the actual dataset, the PCA is then used to clean and rearrange it into a new data structure. Third, WSVM is used to predict stock exchange points using converted data. Finally, they ran a series of tests involving PCA-WSVM, WSVM, PCA-ANN, and a stockbroker from two very popular Chinese markets, 'Shanghai and Shenzhen', to assess the effectiveness of our existing assets.

Yang et al. [22] created a two-step mixed stock option strategy by combining stock forecasts with later stock market options (including stock forecasts and stock options). (1) Using computing intelligence (CI) or a high-quality learning machine with high learning ability and fast computer speed, predict future stock trading. (2) The stock exchange system is designed as a mix of predictable line items (created in the initial stage), valuables (famous in existing books) depending on CI-based measurement, with stocks listed above the weighted portfolio. Using the Chinese share market as a research sample, the artistic results proven that novel's hygienic way, which uses great predictable materials, outperforms conventional methods (excluding stock forecasts), and same counterparts in outcomes. Research proposes a model for stock option selection based on a combination of stock forecasts to better illustrate the future features of the complex stock market. The novel model adds more to the text in two ways. For starters, it would be the first attempt to combine stock forecasts with stock options in order to create a new stock option with predictable features in order to capture upcoming market structures. Second, its tested in the Chinese stock market against traditional stock options models (other than stock forecasts) and comparative partners (along with other forecasting models, design features, development algorithms, and robust functions) to ensure its efficiency and effectiveness.

Sun et al. [20] propose an ARMA-GARCH-NN machine learning method to capture internal trends and predict stock market shocks in this study. To ensure diversity, they combine basic financial pricing models with sensory networks, as well as explicit methods and processes, to analyze high-frequency stock market data in the United States. The findings provide the first indication of market shock. They also ensure that 'ARMA-GARCH-NN' works properly by detecting styles in large datasets without relying on robust distribution estimates. Their method effectively combines the benefits of traditional financial approaches with 'data-driven' methods to uncover masked patterns in massive amounts of financial data. They propose and implement ARMA-GARCH-NN, a data-driven method, in this paper to solve the complex problem of predicting market shocks. In line with this proposed approach, we are looking into the role of past data in detecting unexpected market movements, which we call the ARMA-GARCH model. Furthermore, we propose a near-k verification method for selecting a feature. Their findings suggest that: I IARMA-GARCH market shocks can be predicted internally; (ii) using the MRMR feature and proximity-k proven will result in better consistent efficacy as compared to other measurements (iii) markets volatile movements may be recent trading signs which are helpful in investment decision-making process. Second, it is tested in the Chinese A stock market against traditional stock options models (excluding stock forecasts) and their comparative counterparts (along with other forecasting models, design features, development algorithms, and robust functions) to ensure its efficiency and effectiveness.

Yan & Aasma [25] propose a novel approach to learning prediction, and we have developed a comprehensive model based on stock-CEEMD-PCA-LSTM research to match. The integration of the integration mode (CEEMD), as a sequence of diminishing and diminished modules, can be used in this model to differentiate the variance or trend ratios of different time series, yielding a number of internal mode functions (IMFs) of varying sizes. PCA reduces the size of the IMF component, eliminates external data, and increases response speed by storing more information on raw data. Following that, high-quality features that were not included are separately included in LSTM networks to predict the closing value of each item on the following trading day. Finally, the final predicted value is determined by the sum of the predicted values for each item.

J. Long et al. [11] present an in-depth neural network model that predicts stock price movements based on idle records as well as open data in this paper. Because they rely on other stocks, their method employs a graph of information and methods for embedding the graph in order to select the best target stock for market structure n information about trading. Because of the large count of depositors and the complexations of the job record data, investors are mobilized to lower size in matric trademarks, and matriculants are linked to a convolutional neural network to find investment styles. Finally, the short-term memory network used to track attention can forecast stock price patterns, which can help with financial decisions. Price and style indicator analysis produces far superior results than other forecasting bases. The practice's prediction accuracy has reached 70% or higher, far greater than current Accuracy which is 65%. Outcomes also revealed model's durability and performance.

W. Long, Lu, and Cui [12] propose a 'Multi-Filters Neural Network (MFNN)', a single-type model for excluding features of financial time series and price forecasting functions. By combining convolutional and duplicate neurons, multiple filter formats are created, allowing information to be available from a variety of sources and market ideas. Our MFNN, which is based on the CSI 300 of the Chinese stock market, is used in advanced market forecasts and simulation functions. The positive prediction result from MFNN was 11.47 percent for the best machine learning method and 19.75 percent for the best mathematical method. In comparison to RNN and CNN, the combination of filters and deliberate network design increased accuracy by 7.19% and 6.28%, respectively. Models with deeper learning styles outperform all other foundations in terms of market imitation. Following that, our proposed MFNN outperformed typical machine learning results by 15.41 percent and the mathematical method by 22.41 percent.

Ma, Han, and Wang [14] summaries their predictable impacts on the development of a variety of measurement methods, including variance (MV) and omega portfolio performance, in this study. Portfolio models with integrated return values are used as measurements to demonstrate the length of this variation. The research is carried on 9 years of previous data from 2007 to 2015 of a 100% index of Chinese stocks. The results of the tests show that the MV and omega models have a much higher RF return rate than the other models. Furthermore, the quantity of RF and MVF is greater than the quantity of RF and OF. Because these two types are highly profitable, this paper assesses their effectiveness after deducting the amount of work determined by profit. According to the results of the tests, the RF + MVF model outperforms the MVF model, while the omega model with SVR (SVR + OF) outperforms the models. Furthermore, RF + MVF outperforms SVR + OF, with a higher return on nearly half of all revenue, RF + OF and RF + MVF outperform SVR + OF. As a result, the paper advises investors to create an MVF with an RF return on day-to-day investments.

Seong and Nam [19] proposed a K-means integration method and a multi-learning kernel for predicting stock price increases by combining data from a specific company and its linked collections. They hope to justify outcomes of presented approach to previous alternatives using three years of data from the' Republic of Korea'. The outputs help us to prove the proposed method is more speculative than the available methods in many cases. The findings clearly show that the need for a group analysis is dependent on the classification, and that the total number of collections analysis becomes more important as the diversity increases. They learned how to better integrate company stock prices with companies that could improve performance, broaden the range of impacts on a group with similar patterns in each company in this study. They look for companies with similar stock price movements and combine them to form an equal group to find teams with similar patterns. The GICS programme includes Target, Food Expenses, and Pharmacy. But requirement is that you choose categories which has varying degrees of heterogeneity. The costs of materials, Central Pharmacy, and food vary greatly.

Ismail et al. [8] used a mixed strategy that included machine learning methods and continuous homology to improve prediction. 'Kuala Lumpur Stock Exchange' quotes were chosen from the 'Kuala Lumpur Composite Index'. Indeed, as a result of the return of these stock prices through continuous programming using mechanical learning methods such as retrieval, sensory processing network, supporting vector, and random forest forecasts, continuous homology was used to discover new and useful vectors in fixed topological features. Index of the Kuala Lumpur Composite. The proposed method is compared to others in which machine learning methods are used independently in stock recovery and technical indicators for comparison. Observations the parts connected to the holes are the fixed geographical features used in this study.

Bustos and Pomares-Quimbaya [1] hope to bridge the gap by providing a thorough examination of stock market prediction methods, including segmentation, implementation, and comparison. The experiment focuses on research that forecasts stock market movement from 2014 to 2018, utilising scientific data from Scopus and Web Science. At the same time, it examines research and other updates from newly published studies on the same visual cues. This is a revised version of the stock market forecast. It focuses on activities that were a source of contention in predicting the stock market between 2014 and 2018. In this case, 52 studies were found, with detailed explanations provided depending on the sort of inputs and results used n matching algorithm employed. 'Technical Indicators' are widely used source of information in predicting financial markets also have been shown to be highly speculative data.

In this paper, Thakkar and Chaudhari [21] investigate and familiarise themselves with the number and demand of DNNs in the stock market; investigate the use of DNN variants in temporary stock exchange data; and broaden our research to include hybrids, metaheuristic methods, and DNN methods. Using various DNNs, they obtained potential stock market predictions. They analysed the impacts of such types in predicting markets data using a series of stock market predictors with nine research-based models. Along with it they tested effectiveness of various approaches by varying numbers of aspects. They discussed the concerns and guidelines for upcoming study before concluding our research study. This study could be expanded to include a thorough examination of the DNN-based stock market forecast, which includes the majority of the surveys conducted between 2017 and 2020.

Jing, Wu, and Wang [10] introduced a mixed model in this study that combines an in-depth learning strategy with a stock price analysis model. They use the Convolutional Neural Network model to differentiate between investors' hidden feelings and those expressed in large trading. They then developed mixed research way that utilized 'Long Short-Term Memory (LSTM)' approach of the 'Neural Network' to analyse 'Technical Indicators' from market as well as the results of the previous phase's emotional analysis. Furthermore the project conducted real time test on the Shanghai stock Exchange (SSE) from six three time factories to ensure the proposed models efficiency and effectiveness. The results also proves that the presented study surpasses the basics in classifying investor's sentiments, and that this mixed method outperforms individual model and models that do not consider emotions in predicting stock prices.

Jiang [9] focuses on the most recent advancements in a deep stock market research system rather than providing a comprehensive picture of significant history. Although many studies have shown that in-depth research is an effective way to forecast stock markets, this study doesn't intend to include thorough comparative study in deep learning n forecasting strategies, that may necessitate a large number of accounting resources and may impede future studies.

Haq et al. [6] predict future price movements using an in-depth production model that combines selected features and multiple selection processes to produce a subset of the appropriate feature. First, they used a construction model, vector support machine, and random jungle to calculate the value of an extended set of 44 'Technical Indicators' drawn by 88 daily stock data points. Low level featrues are pared down to their bare essentials, and some are grouped together. The variable value is also used to choose important item within each collection to create end set. Inputs are generated by in-depth production model with market signal indicator and a focus method is used.

Yun, Yoon, and Won [23] used 'GA-XGBoost' mixed program with an modern engineering process that includes, data development, feature extension and better feature set selection. The purpose of this research is to demonstrate the significance of the Engineering in stock price estimates contrasting sets of data acquisitions with real data and optimising prediction performance on measurement models. The inclusion of 67 technical indicators in historical data appears to result in a significant increase in accuracy.

Rezaei, Faaljou, and Mansourfar [18] propose a strategy based on the premise that by combining these species, certain algorithms developed among them can improve the model's analytical capacity. This assumption is supported by empirical evidence, which demonstrates that combining CNN with LSTM, CEEMD, or EMD can improve predictive accuracy and outperform other partners. Furthermore, when compared to EMD, the proposed method with CEEMD outperforms it. The proposed algorithm's basic idea was to integrate CEEMD, CNN, and LSTM models to establish interactions between them, which could result in richer features and timelines. The stock price was then predicted using an expert algorithm.

Hogenboom, Brojba-Micu, and Frasincar [7] present a high-quality traditional language event-based value line that includes word distortions in the event acquisition process. We examine what occurs in environmental news messages and their historical impact on stock prices. Using historical prices and news items obtained from the Dow Jones Newswires over a two-year period, we examine the accuracy of the phrase in price indications in the two NASDAQ-100 business scenarios. They evaluate the buy and sell signals' accuracy based on our predicted stock price movements, as well as the excess return provided by the trading strategy that generates these signals. Over the next two days, event-based price estimates appear to be very reliable. When a name-based system variation changes the process of measuring the size of an event acquisition process, the number of events received typically drops by more than 30%, greatly reducing the noise presented in confusing event stock estimates. As a result, even minor improvements in the accuracy of buying and selling signals generated based on these forecasts frequently result in significant improvements in nearly 70% of the additional benefits associated with the average. They developed and tested a high-quality event-based natural language pipeline analysis in this paper, allowing WSD to be included in the pricing process for key event events. It is used to identify events in environmental news messages and compare their historical impact on stock prices.

Lu et al. [13] employ a technique that combines Convolutional neural networks (CNN), Bi-directional Long-term Memory (BiLSTM), and attention-grabbing (AM) techniques. CNN is a feature extraction algorithm that is used to extract features from input data. Bi-LSTM forecasts stock closing rates for the next day based on the feature data obtained. AM is used to capture the beneficial effect of the stock closing price on various previous periods in order to improve forecast accuracy. To demonstrate the efficacy of this strategy, 1000 Shanghai Composite Index trading days were used to forecast the next day's stock closing using this method and seven others. The results show that the method's effectiveness has improved, with MAE and RMSE at very low levels (at 21.952 and 31.694). R2 is the greatest number (its value is 0.9804). The CNN-BiLSTM-AM method accurately predicts stock prices and provides a more reliable platform for investors to make stock investment decisions when compared to other methods.

**Table 1:** Overview of recent research in financial market prediction

| citation | Title | Methodology/ approaches | Dataset/Market | Performance measure | Advantages/Pros | Limitations/cons |
|---|---|---|---|---|---|---|
| Patel et al. [16] | Predicting stock and stock price index movement using Trend Deterministic Data Preparation and machine learning techniques | Four prediction models ANN, SVM, RF, NB | Stocks- Infosys, reliance<br><br>Indices- 'CNX Nifty and S&P Bombay Stock Exchange (BSE) Sensex'<br>Year- 2003-2012 | Accuracy, F-measure | Improved performance when models evaluated with trend deterministic data | Only focuses on short term prediction<br><br>Poor performance when technical parameters used as continuous values |
| Patel et al. [17] | Predicting stock market index using | Two stage fusion approach (SVR– | Indices- 'CNX Nifty and S&P Bombay | MAPE, MAE rRMSE, MSE | mixed model performs effectively as compared | No real time prediction |
| | fusion of machine learning techniques | ANN, SVR–RF and SVR–SVR) | Stock Exchange (BSE) Sensex' | | to individual forecasting approaches<br><br>Helps to set stop loss order | Doesn't consider national, global financial news |
| Y. Chen and Hao [3] | Integrating principle component analysis and weighted support vector machine for stock trading signals prediction | PCA, WSVM | 'Shanghai and Shenzhen stock exchange markets' | Accuracy, Profit percentage | In real-world applications, it is reliable and which can be used to forecast buy and sell signals.<br><br>Stocks in selloffs, consistent trends, and uptrends have superior precision. | Doesn't consider different investing strategies which might give mediocre profits even if high accuracy of the prediction |
| Nam and Seong [15] | Financial news-based stock movement prediction using causality analysis of influence in the Korean stock market | Multiple kernel learning, Transfer Entropy and Causal relationship | Korean market | Average, Standard Deviation | Market values are forecasted even if there is no business news related to chosen stock | Doesn't consider historic data only dependent on financial news |
| Sun et al. [20] | Exploiting intra-day patterns for market shock prediction: A machine learning approach | ARMA-GARCH-NN | U.S. stock market | Accuracy, RMSE | Can capture unexpected movement of market | Unexpected changes which occurs nationally or globally have big impact on forecasting |
| Y. Chen and Hao [4] | A feature weighted support vector machine and K-nearest neighbor algorithm for stock market indices prediction | FWSVN and KNN | Chinese stock market indices- Shanghai and Shenzhen | RMSE, MAPE | Improved short, medium, and long-term forecasting correctness | Good prediction accuracy on Chinese stock market only<br><br>Short and medium term have lower prediction accuracy compared to long term |

| W. Chen et al. [2] | Mean–variance portfolio optimization using machine learning-based stock price prediction | Mean–variance model, eXtreme Gradient Boosting Firefly Algorithm | Shanghai Stock Exchange | MAPE, MSE, MAE, RMSE | Superior than traditional approaches when risks and return–risk ratio are to be considered | only Shanghai market facts were chosen for study |
|---|---|---|---|---|---|---|
| X. dan Zhang, Li, and Pan [24] | Stock trend prediction based on a new status box method and AdaBoost probabilistic support vector machine | status box method, AdaBoost, probabilistic support vector machine | 'Shenzhen Stock Exchange' (SZSE) 'National Association of Securities Dealers Automated Quotations' (NASDAQ) | g-means, Accuracy | Long term turning trend can be easily detected | Problem in predicting short term trends<br><br>Real time prediction problem |
| Rezaei, Faaljou, and Mansourfar [18] | Stock price prediction using deep learning and frequency decomposition | EMD-CNN-LSTM CEEMD-CNN-LSTM | DAX, and Nikkei225, S&P500, Dow Jones<br><br>Year- from January 2010 till September 2019 | RMSE, MAE, MAPE | Better prediction accuracy as compared to traditional Hybrid ML models | Multi-Layer CNN and LSTM models not verified |

## 3. Research Dataset

Reliance and Infosys are the stocks of NSE (National Stock Exchange) are selected for the study. Historical prices of total ten years from (Jan 2012 to Dec 2021) is practiced for this approach. We have used Yahoo Finance <https://finance.yahoo.com/> website to obtain whole research data which is required for the study. The head snippets of each are as given below.

**Table 2:** Reliance Stock Snippet

| Date | Open | High | Low | Close | Adj Close | Volume |
|---|---|---|---|---|---|---|
| 2021-12-24 | 2370.000000 | 2392.000000 | 2337.550049 | 2372.800049 | 2372.800049 | 3639616.0 |
| 2021-12-27 | 2361.550049 | 2378.000000 | 2348.100098 | 2370.250000 | 2370.250000 | 1853948.0 |
| 2021-12-28 | 2375.600098 | 2404.850098 | 2373.050049 | 2398.399902 | 2398.399902 | 2941883.0 |
| 2021-12-29 | 2391.000000 | 2419.000000 | 2382.100098 | 2402.500000 | 2402.500000 | 7118779.0 |
| 2021-12-30 | 2400.000000 | 2404.949951 | 2345.600098 | 2359.100098 | 2359.100098 | 13537254.0 |

**Table 3:** Infosys Stock Snippet

| Date | Open | High | Low | Close | Adj Close | Volume |
| --- | --- | --- | --- | --- | --- | --- |
| 2021-12-24 | 1872.949951 | 1875.750000 | 1854.000000 | 1863.500000 | 1863.500000 | 3780347.0 |
| 2021-12-27 | 1860.000000 | 1874.500000 | 1845.050049 | 1866.150024 | 1866.150024 | 2647733.0 |
| 2021-12-28 | 1880.699951 | 1895.900024 | 1878.400024 | 1888.000000 | 1888.000000 | 3340933.0 |
| 2021-12-29 | 1883.500000 | 1893.800049 | 1876.400024 | 1885.550049 | 1885.550049 | 3236635.0 |
| 2021-12-30 | 1884.500000 | 1909.800049 | 1874.349976 | 1892.849976 | 1892.849976 | 4584738.0 |

The dataset represents daily OHLC means (Open High Low Close) Adj Close and volume prices. Some dataset also represents the stock data in the form of OHLC diagram as showed below, which help us to know (Open High Low Close) in selected time frame. This diagrams are helpful because it shows us primary data throughout time, where closing price is said to be more relevant by most of the researchers.

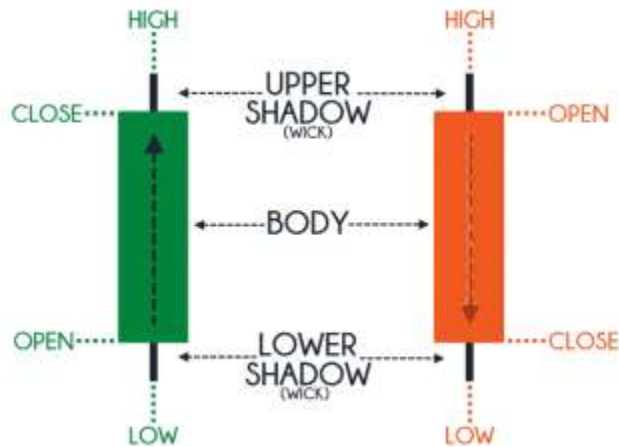

**Fig. 1** Candlestick Diagram

Let's understand what all these value means:

- Open Price: Open means the price at which a stock started trading for a given time frame.
- High price: Maximum price the stock for given time frame.
- Low Price: For a certain time window, the stocks minimum price.
- Closing Price: Closing price refers to value of single stock at completion of specific timeframe.

- Adj. Close: The daily closing price tells only about the cash component of a stock. It is the price at which the last lot of the stock is bought or sold in the last trading session. The adjusted closing price takes into account all the factors that may have affected the stock price after the market hours which includes corporate events such as dividends, stock splits and rights offerings are factored by Adj. closing price.
- Volume: Volume basically is the number which will tell us the number of share traded (buy and sell) in a chosen timeframe (usually daily). Whenever any stock related news goes into public or markets changes dramatically, high day volume is common. While less volume is common on a normal uneventful days.

## 4. Methodology and implementation

There are various machine learning and deep learning forecasting model to choose from, LSTM is chosen for this case study.

*4.1 Prediction Model*

There are various approaches presented to solve real world problems. Algorithms which are designed to solve such problems are considered as one of the most challenging one to complete. Such real world problem includes, predicting the weather, understanding the patterns in large amount of data and use it for future prediction, hearing the voice and translating it into text, language translation, predicting the next world while you type in your word file. Various problems are categories on the basis of data and the result we want from those data. It is observed that for most of the time series data and conditions where we require to remember the previous state outputs, LSTM have showed significant improvement in the performance as compared to other models that were designed to do such tasks. It is also observed that because of its ability to remember for a longer period of time, LSTM performs better in sequence prediction problem as compared to the conventional Feed Forward Network (FFN) and Recurrent Neural Network (RNN).

Every LSTM cell has input, output and forget gate. Data that enters the LSTM's network is kept, while the unnecessary data is erased by the forget gate. LSTM can be used in a variety of applications, including weather forecasting, natural language processing (NLP), speech recognition, handwriting recognition, time-series prediction, and so on.

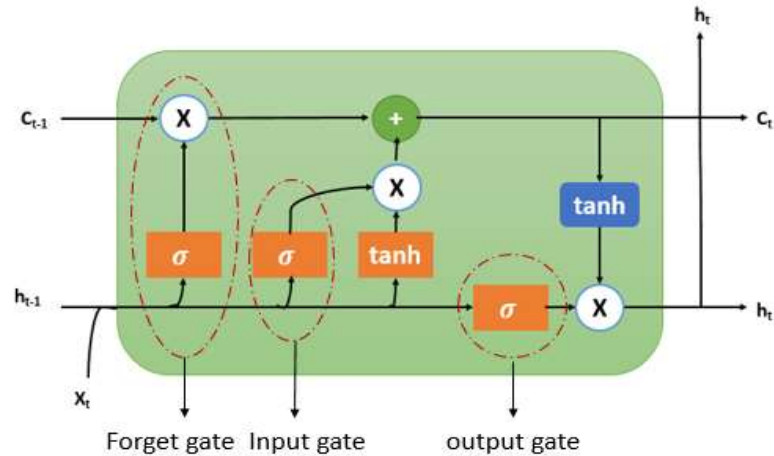

**Fig. 2** LSTM cell

As shown in above Fig. 5 the inputs to the current cell state ($c_t$) is the prior hidden state ($h_{t-1}$), prior cell state ($c_{t-1}$) and present input ($x_t$). Input gate, output gate and forget get make up the one LSTM cell.

**Forget Gate:** It will remove unnecessary data from the cell. By multiplying the hidden state by a sigmoid function, the information which is less important or not required for the LSTM to understand things is removed. This phase is critical for the model's performance to be optimised. It uses two inputs i.e., $h_{(t-1)}$ and $x_t$, where $h_{(t-1)}$ previous cell hidden state output and $x_t$ present cell input as shown in equation (1).

$$f_t = \sigma(w_{fx} * x_t + w_{fh} * h_{t-1} + b_f) \quad \text{------(1)}$$

**Input Gate:** This cell is in charge of controlling the data that is fed into it from the input. To filter some input, a forget get is utilised. Using tanh function a vector is produced by summing all the possible from prior cell hidden state $h_{(t-1)}$ and present cell input $X_t$. tanh function gives the result which ranges between [-1, 1]. Finally, the outputs of sigmoid and tanh functions are multiplied and the output is summed up with cell state. The mathematical formula for it is shown in equation (2).

$$i_t = \sigma(w_{ix} * x_t + w_{hh} * h_{t-1} + b_i) + \tanh(w_{cx} * x_t + w_{ch} * h_{t-1} + b_i) \quad \text{------(2)}$$

**Output Gate:** The Tanh function is used to construct a vector from the cell state with all possible values. Sigmoid function is applied on prior cell hidden state $h_{(t-1)}$ and present cell input $x_t$ to filter necessary data from the previous cell. Now, the outputs of sigmoid and tanh functions are multiplied and this output is sent as a hidden state of the next cell.

$$o_t = \sigma(w_{ox} * x_t + w_{hh} * h_{t-1} + w_{oc} * c_{t-1} + b_i) \quad \text{------(3)}$$

We get Intermediate cell state ($c_t$) by multiplying forget gate ($f_t$) with prior cell state ($c_{t-1}$). Then this intermediate state is added to the output of the input gate as hsown in equation (4).

$$c_t = f_t * c_{t-1} + i_t \quad \text{------(4)}$$

Current hidden/output state is obtained by multiplying output gate and tanh of cell state.

$$h_t = o_t * \tanh(c_t) \quad \text{------(5)}$$

*Univariate Time Series:* This is primarily concerned with a single variable. The hypothesis here is that the value of a time - series data at time step 't' is derived essentially from the earlier time steps t-1, t-2, t-3, and so forth. These models are less difficult to create than multivariate models. The variable of interest in stock price forecasting is usually the closing price of a capital asset.

*Multivariate Time Series:* Here, we don't have single dependent variable unlike univariate. We can have many interdependent variables. It can take many influencing factors into account but other several non-influencing factors can also be considered. Multivariate stock market forecasting models are not only dependant on 'closing prices' but many other variables like opening price, daily highs, volume, moving averages and so on. This models are generally more complicated to develop when compared to the univariate models.

*4.2 Technical Indicators*

There are a number of technical indicators that can be applied to forecast future stock movement. This analysis employs a total of ten technical indicators. Table 4 shows the top ten technical indicators we used in our research.

**Table 4:** Formulas for Technical Indicators [16]

| Indicators Name | Formula |
|---|---|
| Simple Moving Average (10) | $\dfrac{C_1 + C_2 + \cdots + C_{10}}{10}$ |
| Simple Moving Average (50) | $\dfrac{C_1 + C_2 + \cdots + C_{50}}{50}$ |
| Weighted Moving Average (10) | $\dfrac{10C_1 + 9C_2 + \cdots + C_{10}}{10 + 9 + \cdots + 1}$ |
| Exponential Moving Average(EMA) | $Price(t_d) \times k + EMA(y_d) \times (1-k)$ |
| RSI | $100 - \left[\dfrac{100}{1 + \dfrac{Avg\ Gain}{Avg\ Loss}}\right]$ |
| CCI | $\dfrac{M_t - SM_t}{0.015 Dt}$ |
| AD (Accumulation Distribution) | $\dfrac{H_t - C_{t-1}}{H_t - L_t}$ |
| Stochastic K% | $\dfrac{C_t - LL_{t-(n-1)}}{HH_{t-(n-1)} - LL_{t-(n-1)}} X 100$ |
| Stochastic D% | $\sum_{i=0}^{n-1} \dfrac{k_{t-i}}{10} \%$ |
| MACD | $\dfrac{MACD(n)_{t-1} + 2}{(n+1)x(DIFF_t - MACD(n)_{t-1})}$ |

Where, C-closing value, L- low value, H- high value, $t_d$- today, $y_d$- yesterday, $H_t$- highest price, $C_t$ – closing price at time t, $DIFF_t$=EMA (12)$_t$-EMA (26)$_t$, EMA exponential moving average, $LL_t$ and $HH_t$ implies lowest low and highest high in the last t days, k is the time period of k-day exponential moving average, N=number of days in EMA k=2÷(N+1), $M_t = \dfrac{H_t + L_t + C_t}{3}$, $SM_t = \dfrac{\left(\sum_{i=1}^{n} M_{t-i+1}\right)}{n}$, $D_t = \dfrac{\left(\sum_{i=1}^{n} |M_{t-i+1} - SMt|\right)}{n}$

**SMA** is obtained by multiplying recent values to no of times the calculation measure has been employed. As the starting point data, first ten day average values are calculated then initial price will be reduced in the next data point, then increased by day 11 before taking a rate hence the same is applied furthermore. To get SMA (50) on the other hand, can collect enough data to keep a consistent average over 50 days. The simple moving average shifts the volatility. SMA pointing upwards indicates the price of stock is rising. If it's pointing down, it means the mortgage rate is going down.

*EMA* which offers lot of value and significance for recent prices. At the moment, the explicit moving average reacts more strongly to recent price changes than the moving average with the same weight in all observations. The EMA is a dynamic scale that gives a lot of weight and importance to the most recent data points. Similar to other moving averages this also help us to get purchase or sell signals. It frequently employs a variety of EMA durations, such as 10 days, 50 days, and 200 days.

*RSI* is a calculated by measuring the size of recent price movements, a technical analysis speed indicator assesses the conditions of the most bought or sold stock. The index was created by 'J. Welles Wilder Jr.' which was later published in a book, "New Ideas in Technology Trading Systems," in 1978. It is represented as a line graph between two extremes 0 and 100. The RSI provides technical traders with signal of bullish and bearish price pressures.

*CCI* was created to identify long-term fashion trends, but retailers have adapted it for use in all markets and at all times for aggressive traders, frequent trading generates buy and sell signals. Traders frequently use the CCI to determine the long-term trend and to identify pullbacks and generate trading signals on the short-term chart. When compared to other indicators, stochastics is a popular technical indicator because it is easy to use and offers a high level of precision. It is a type of technical indicator known as an oscillator. Traders can enter or exit positions based on the index's dynamics by receiving buy and sell signals. Stochastics is used to determine when a stock has been undervalued or overvalued.

*AD* indicator takes the help of volume and prices to know if stock price is accumulating or dispersing. This one generally focuses on the difference between the stock prices and volume flow. As a result, if stock price is increasing and the indicator is giving us downward signal that means buyers volume is very low to sustain the raise and hence as a result the price will fall down in future.

*Stochastic oscillator,* a renowned momentum indicator which help us to know if stock is undervalued or overvalued. To generate the signal, it compares closing values and range of values in given timeframe.

*MACD* is a trend that follows a dynamic index that demonstrates the relationship between the moving average values of two securities. It is basically the difference between $EMA_{(26)}$ and $EMA_{(12)}$. Technical signals are formed when indicator crosses upwards or downwards the signal line. The MACD can help investors to determine whether a price movement is bullish or bearish.

### 4.3. Prediction Approaches

The input data for this study is gathered in two ways. In this study, both approaches are investigated.

*4.3.1 Close Value Approach-* In this approach, we fed closing prices directly to the prediction model and validated its performance and price behaviours.

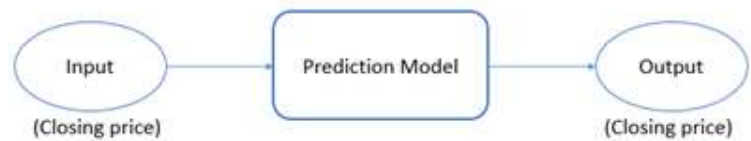

**Fig. 3:** Close Value Approach

Summary statistics of the input data used in this approach is as follows:

**Table 5:** Reliance

|       | Open        | High        | Low         | Close       | Adj Close   | Volume       |
|-------|-------------|-------------|-------------|-------------|-------------|--------------|
| Count | 2464.000000 | 2464.000000 | 2464.000000 | 2464.000000 | 2464.000000 | 2.464000e+03 |
| Mean  | 927.099427  | 937.645912  | 915.722374  | 926.149327  | 909.927121  | 8.666866e+06 |
| Std   | 625.053790  | 631.967063  | 616.629926  | 623.754545  | 630.016952  | 6.170090e+06 |
| Min   | 334.330872  | 338.194244  | 333.365021  | 334.875702  | 309.114899  | 0.000000e+00 |
| 25%   | 440.394257  | 445.960243  | 436.419441  | 440.728584  | 419.241432  | 5.199650e+06 |
| 50%   | 538.644196  | 543.052429  | 531.883270  | 537.108734  | 520.039337  | 6.921992e+06 |
| 75%   | 1269.516266 | 1280.784454 | 1256.266846 | 1266.618744 | 1252.863617 | 9.801447e+06 |
| Max   | 2742.750000 | 2751.350098 | 2708.000000 | 2731.850098 | 2731.850098 | 6.584835e+07 |

**Table 6:** Infosys

|       | Open        | High        | Low         | Close       | Adj Close   | Volume       |
|-------|-------------|-------------|-------------|-------------|-------------|--------------|
| Count | 2464.000000 | 2464.000000 | 2464.000000 | 2464.000000 | 2464.000000 | 2.464000e+03 |
| Mean  | 646.988266  | 653.877734  | 640.074252  | 646.946860  | 589.382418  | 8.941005e+06 |
| Std   | 348.833370  | 352.254027  | 345.539891  | 348.903792  | 365.744094  | 8.416750e+06 |
| Min   | 270.000000  | 270.500000  | 257.568756  | 265.475006  | 211.840622  | 0.000000e+00 |
| 25%   | 447.496872  | 450.081246  | 442.789062  | 446.615624  | 370.656746  | 5.441703e+06 |
| 50%   | 542.399994  | 547.500000  | 534.162506  | 541.637512  | 4620543107  | 7.128783e+06 |
| 75%   | 720.000000  | 727.900024  | 712.149979  | 719.837494  | 672.829620  | 9.920350e+06 |
| Max   | 1884.500000 | 1909.800049 | 1878.400024 | 1892.849976 | 1892.849976 | 1.663204e+08 |

*4.3.2 Technical Indicators Approach-* Close Values are used to construct ten technical indicators, which are then used to feed as a input to prediction models. All technical indicators' values are adjusted in range of [1, +1], hence a higher value of any indicator doesn't overshadow a lower value. For this representation of inputs, performance of the model is evaluated.

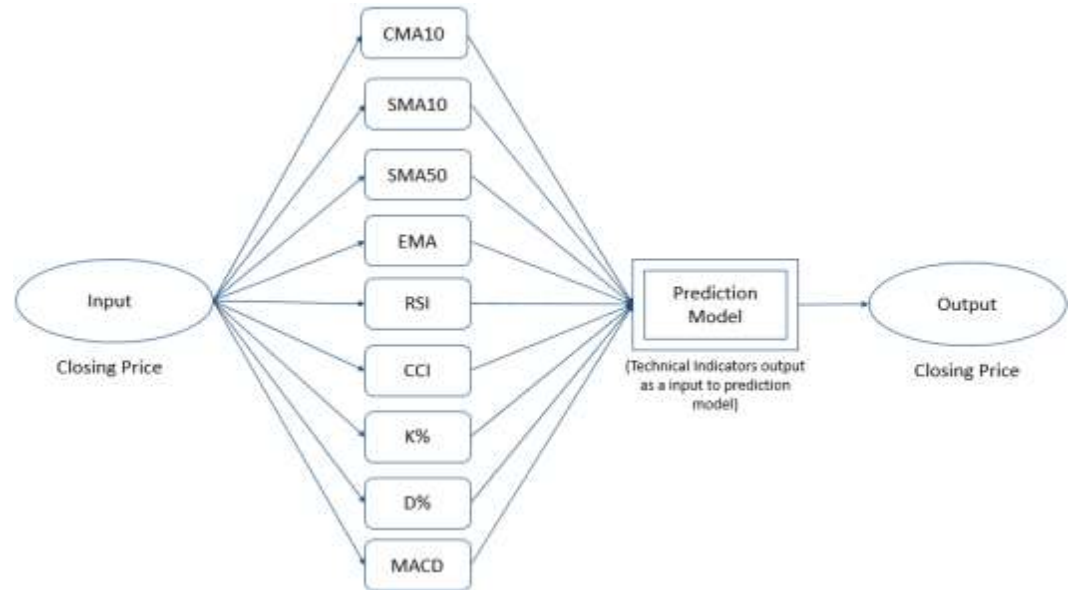

**Fig. 4:** Technical Indicators Based Approach

Summary statistics of the input data used in this approach-

**Table 7:** Reliance

|  | Close | CMA | SMA10 | SMA50 | SMA200 | EMA_0.1 | RSI | K% | D% | CCI | macd | macd s | macd_h |
|---|---|---|---|---|---|---|---|---|---|---|---|---|---|
| count | 2449.00 | 2449.00 | 2449.00 | 2449.00 | 2449.00 | 2449.00 | 2449.00 | 2449.00 | 2449.00 | 2449.00 | 2449.00 | 2449.00 | 2449.00 |
| mean | 929.58 | 524.53 | 925.92 | 909.17 | 849.51 | 922.16 | 52.97 | 53.25 | 53.27 | 12.24 | 5.89 | 5.94 | -0.05 |
| std | 624.10 | 151.28 | 620.89 | 605.15 | 546.59 | 616.97 | 11.71 | 29.87 | 27.81 | 108.21 | 24.35 | 23.01 | 7.00 |
| min | 334.87 | 365.54 | 340.40 | 352.77 | 365.54 | 346.61 | 16.99 | 1.38 | 2.51 | -304.81 | -131.44 | -112.44 | -32.91 |
| 25% | 441.58 | 415.57 | 440.84 | 436.28 | 448.42 | 437.43 | 44.53 | 26.75 | 27.87 | -70.41 | -4.63 | -4.23 | -2.60 |
| 50% | 539.36 | 447.02 | 538.61 | 521.55 | 505.81 | 531.04 | 52.38 | 55.50 | 55.74 | 17.76 | 2.36 | 2.29 | -0.00 |
| 75% | 1267.53 | 619.59 | 1262.68 | 1248.49 | 1215.38 | 1265.31 | 61.38 | 80.48 | 79.05 | 93.88 | 11.33 | 10.90 | 2.49 |
| max | 2731.85 | 926.14 | 2677.67 | 2542.65 | 2243.18 | 2614.91 | 84.45 | 98.85 | 97.53 | 415.84 | 119.20 | 104.77 | 38.65 |

**Table 8:** Infosys

|       | Close   | CMA    | SMA10   | SMA50   | SMA200  | EMA_0.1 | RSI    | K%     | D%     | CCI     | macd   | macd_s | macd_h |
|-------|---------|--------|---------|---------|---------|---------|--------|--------|--------|---------|--------|--------|--------|
| count | 2449.00 | 2449.00| 2449.00 | 2449.00 | 2449.00 | 2449.00 | 2449.00| 2449.00| 2449.00| 2449.00 | 2449.00| 2449.00| 2449.00|
| mean  | 648.81  | 434.91 | 645.98  | 634.42  | 594.43  | 643.38  | 53.53  | 55.66  | 55.63  | 20.97   | 4.06   | 4.01   | 0.05   |
| std   | 349.14  | 84.01  | 345.13  | 329.75  | 271.91  | 341.41  | 11.94  | 28.94  | 26.68  | 108.21  | 12.52  | 11.75  | 3.77   |
| min   | 265.47  | 312.62 | 272.30  | 291.13  | 302.38  | 279.85  | 14.86  | 1.01   | 1.89   | -398.86 | -59.97 | -47.10 | -23.56 |
| 25%   | 449.53  | 354.23 | 449.33  | 440.70  | 419.27  | 452.26  | 45.70  | 30.79  | 32.97  | -60.56  | -2.74  | -2.50  | -1.65  |
| 50%   | 542.47  | 441.05 | 542.45  | 540.33  | 527.74  | 544.59  | 53.60  | 58.76  | 58.47  | 33.04   | 3.52   | 3.34   | 0.19   |
| 75%   | 720.84  | 491.71 | 721.15  | 714.92  | 711.97  | 720.08  | 61.79  | 81.61  | 79.50  | 100.24  | 9.23   | 9.08   | 2.03   |
| max   | 1892.84 | 646.94 | 1850.96 | 1757.72 | 1588.28 | 1818.36 | 84.38  | 99.03  | 97.45  | 446.20  | 58.36  | 54.24  | 16.72  |

## 5. Experimental Result

This study uses a total of ten years of past facts of two highly voluminous stocks, infosys and reliance, from January 2012 to December 2021. These stocks' close prices are used to create the 10 technical indicators. The dataset for both the stock is obtained from <https://finance.yahoo.com/> website.

*5.1 Evaluation measures*

We have calculated error measures to evaluate the viability of our two approaches provided in methodology and implementation section. Mean Absolute Percentage Error (MAPE), Mean Absolute Error (MAE), Mean Squared Error (MSE) and Root Mean Squared Error (RMSE) are used to validate performance of the proposed models. The following equations Equation (6-9) shows their formulas.

$$\text{MAPE} = \frac{1}{n}\sum_{t=1}^{n}\left(\frac{|real_t - predict_t|}{|real_t|}\right) X100 \quad --(6)$$

$$\text{MAE} = \frac{1}{n}\sum_{t=1}^{n}\left(\frac{|real_t - predict_t|}{|real_t|}\right) \quad --(7)$$

$$\text{MSE} = \frac{1}{n}\sum_{t=1}^{1}(real_t - predict_t)^2 \quad --(8)$$

$$\text{RMSE} = \sqrt{\text{MSE}} \quad --(9)$$

We carried out experiment with two approaches hence will see the result for each of the approach mentioned in methodology and implementation section.

*5.2 Approach 1-*

This is the simple approach, here we directly apply the closing prices of selected stocks, Reliance and Infosys on the LSTM prediction model. LSTM model will then try to find out and understands patterns from the previous data. Here, we have used only one attribute from the dataset which is 'closing price', hence the model used is univariate. The model is trained with previous data years of data. The models result are explained in the end of present section.

| Reliance | Infosys |
|---|---|
| 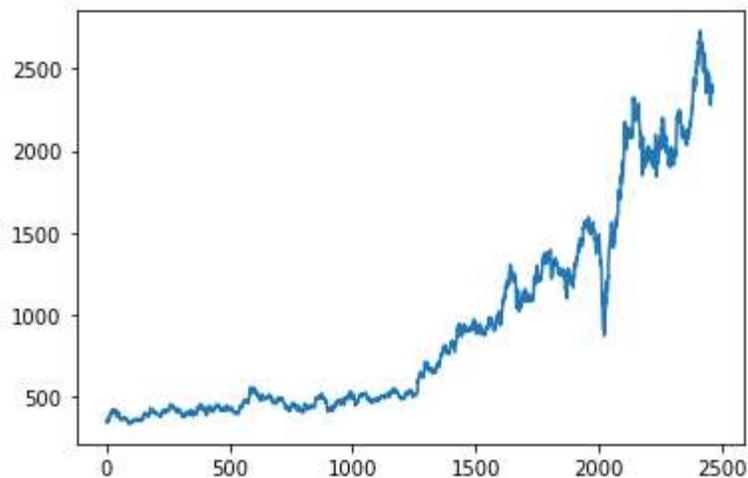 | 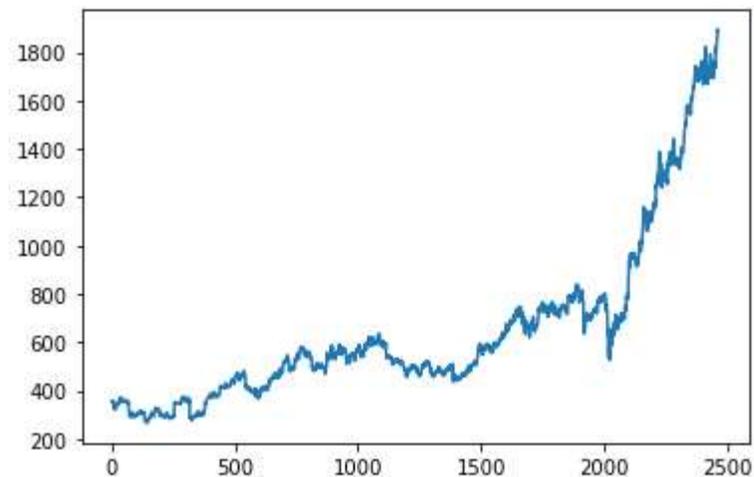 |
| **Fig. 5** Ten years (2003-2012) Closing Prices | **Fig. 10** Ten years (2003-2012) Closing Prices |
| 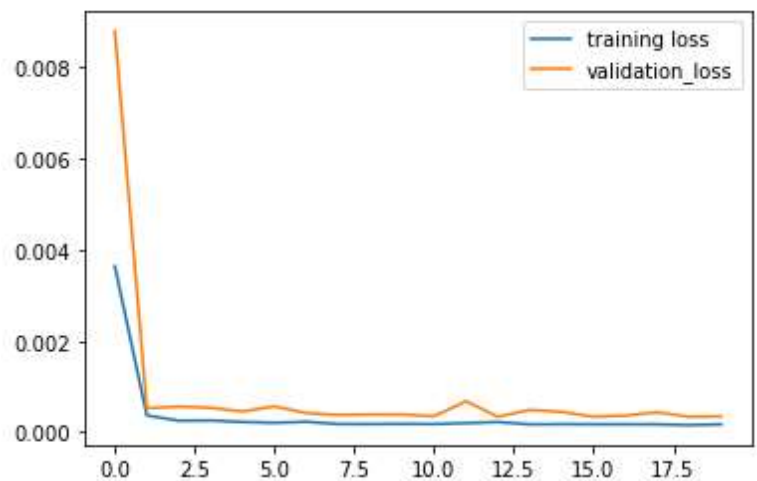 | 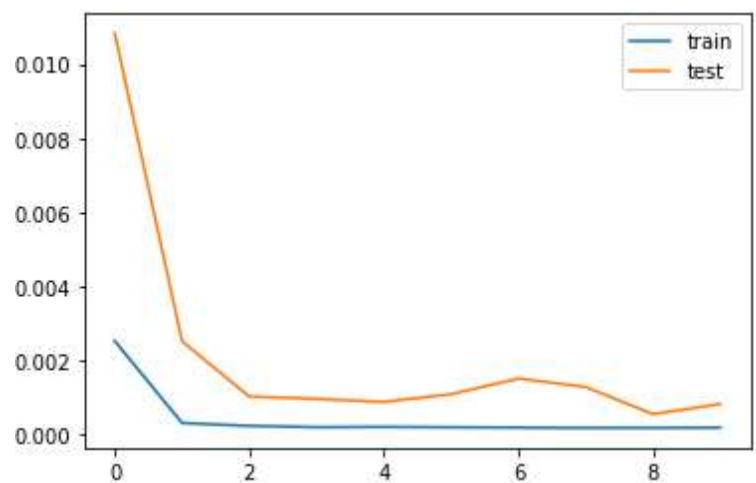 |
| **Fig. 6** Training-Validation loss curve | |

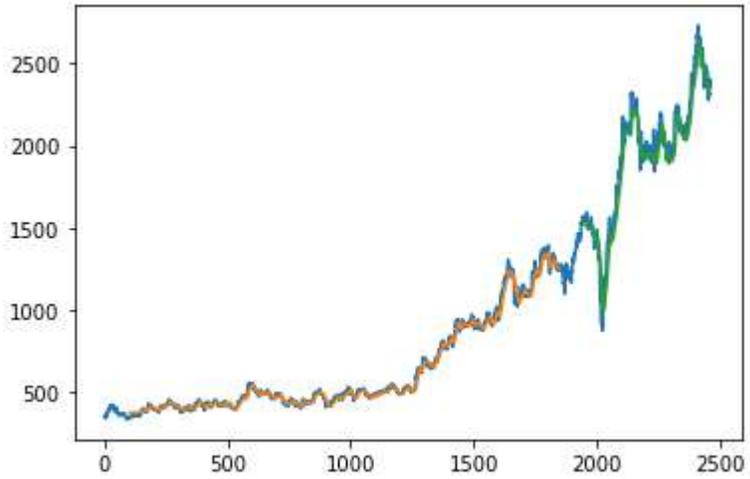

**Fig. 7** Train Test Split

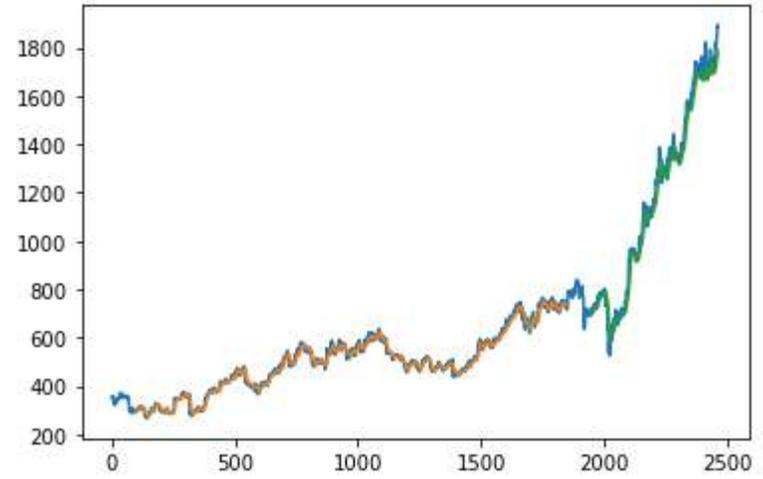

**Fig. 12** Train Test Split

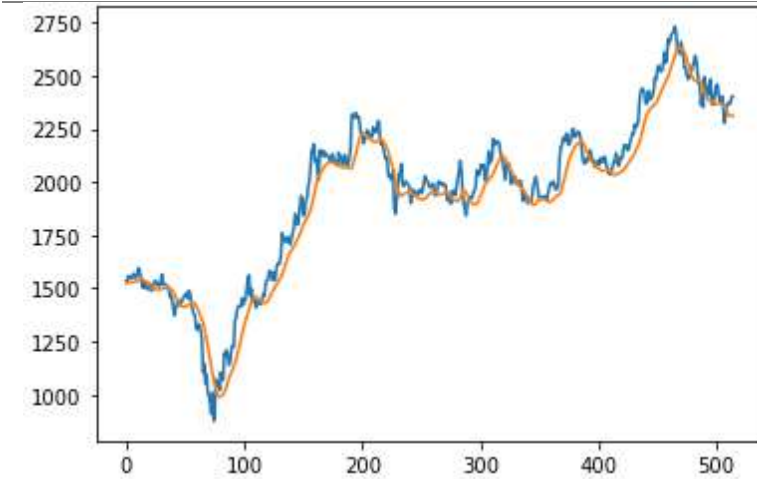

**Fig. 8** Test Data Prediction

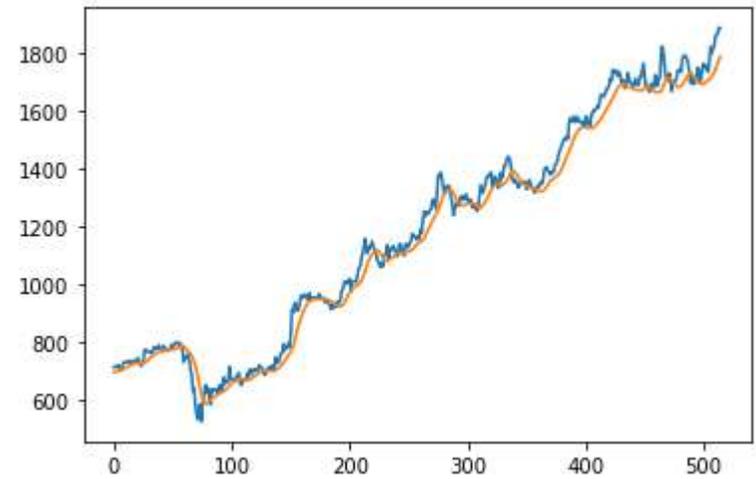

**Fig. 13** Test Data Prediction

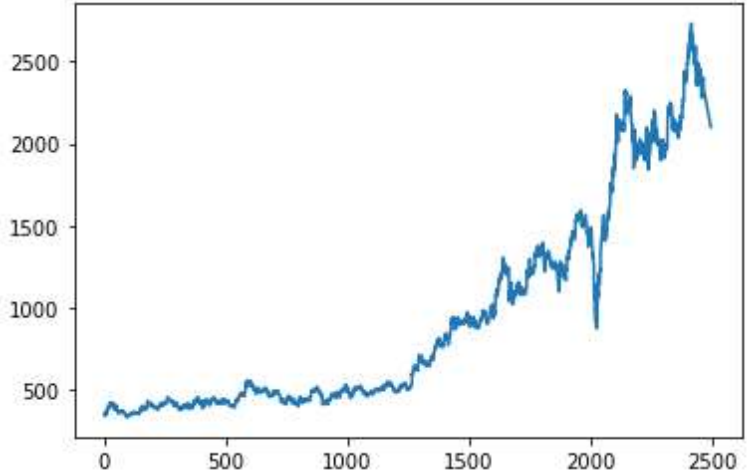

**Fig. 9** 30 Day Prediction

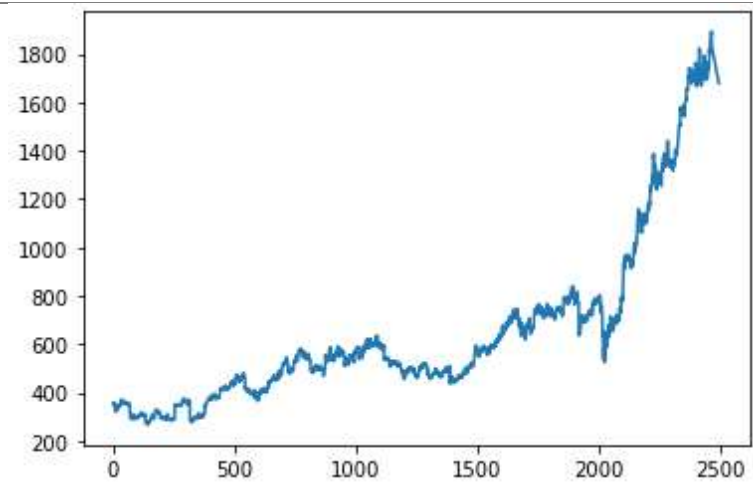

**Fig. 14** 30 Day Prediction

The prediction model under this approach is capable of only predicting the trends. This is a simple approach and hence we are only able to know about the downtrend or uptrend of the stock but not the accurate values as complex prediction model does. The prediction model has performed well and given the downtrend signal for both the stock. Hence, in the Fig. 9 of Reliance and Fig. 14 of Infosys we can see simple small downward line at the end of graph, which basically represents trend, here it is going down means downtrend which means in the upcoming 30 days the prices of the stock will go down. We will not get any clear idea about the price behavior or actual prices for upcoming 30 days. Hence we go for the next approach.

### 5.3 Approach 2

Here, in the second approach unlike in approach 1 we will not consider only the closing prices. We will generate more data with the help of closing prices. As discussed earlier we have chosen possibly the best technical indicators. All the values of technical indicators are calculated as per the formulas mentioned in Table 4. Hence, closing prices along with technical indicators prices are then feed to multivariate LSTM model. Thereafter, model is trained and evaluated for predictions. This methods outcomes are explained further.

| Reliance | | | | | Infosys | | | | |
|---|---|---|---|---|---|---|---|---|---|
| | **Error Measures** | | | | | **Error Measures** | | | |
| | MAPE | MAE | MSE | RMSE | | MAPE | MAE | MSE | RMSE |
| Epochs 20 | 26.67 | 478.38 | 234682.24 | 484.44 | Epochs 40 | 6.24 | 110.56 | 18847.97 | 137.28 |

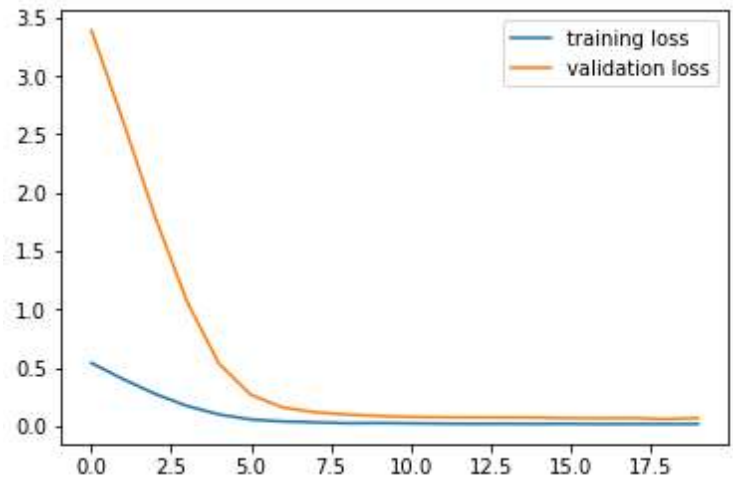
**Fig. 15** Training-Validation loss

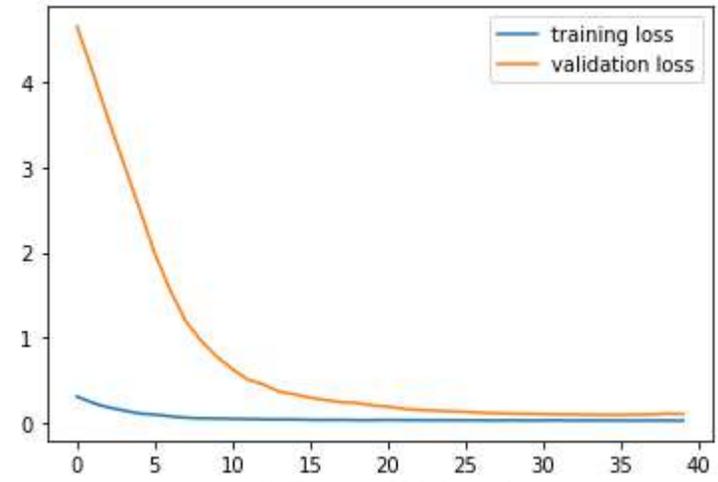
**Fig. 18** Training-Validation loss

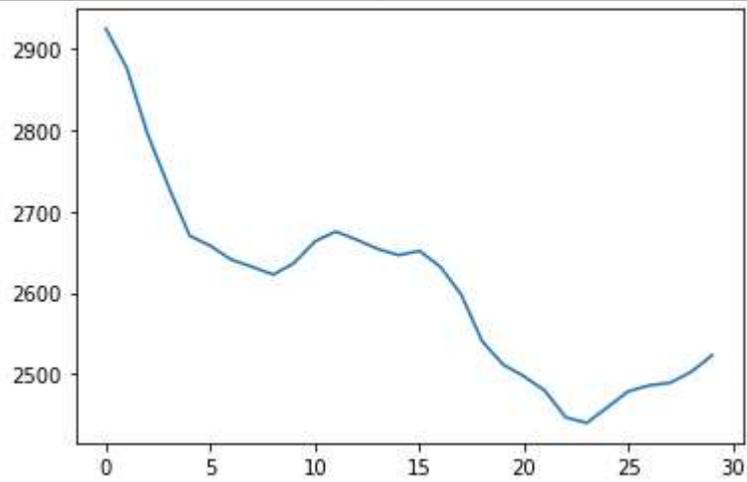
**Fig. 16** 30 days prediction

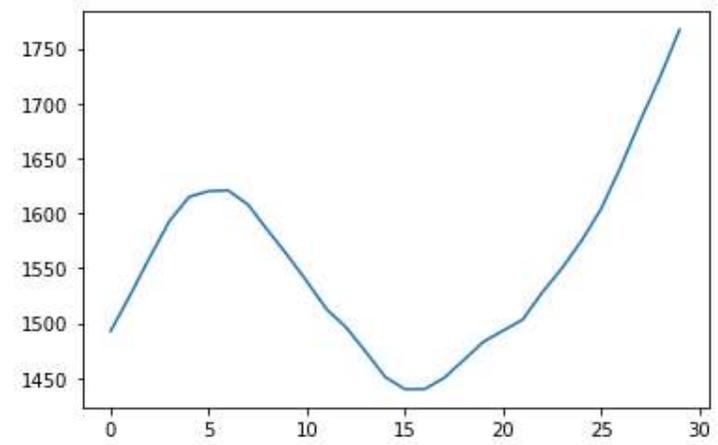
**Fig. 19** 30 days prediction

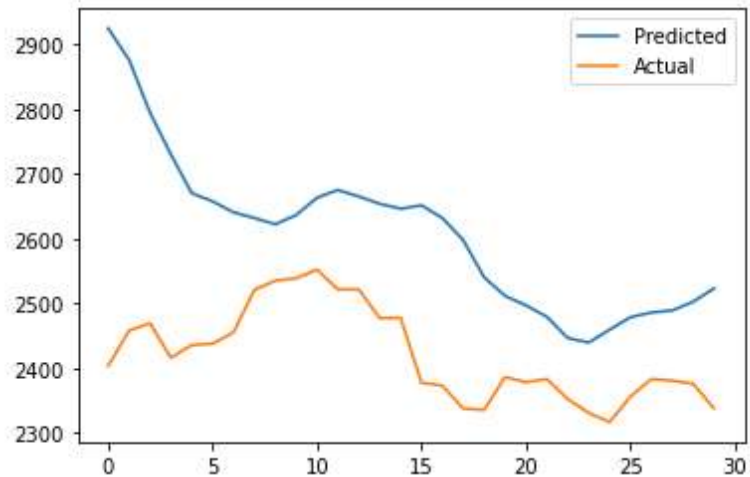
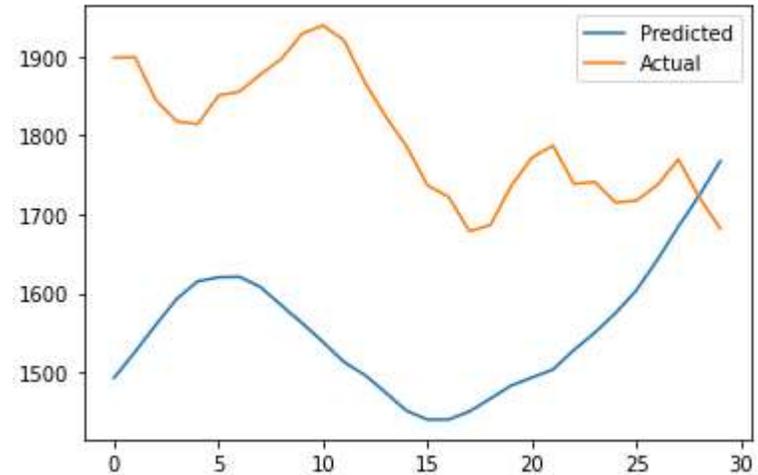

**Fig. 17** Actual vs Predicted

**Fig. 20** Actual vs Predicted

Technical indicators approach along with closing prices performed better as compared to the approach 1 as discussed earlier in 5.2. Here, we are able to get the actual stock values unlike in the approach 1 we are only able to know about the uptrend or downtrend. In fact we can say that approach 1 gives us blurred picture about future price behavior. But, the second approach gives us clear picture about future prices. Fig. 17 & Fig. 20 (Actual vs Predicted) graphs also shows how similar patterns are observed with actual ones. Hence, we can say that technical indicators approach with Multivariate LSTM model will be able to give us idea about future stock price behavior if all the unfavorable conditions are ignored.

## 6. Conclusion

This article's goal is to forecast stock and price trends. Over a ten-year period (2012–2021), performance was predicted using historical data of Infosys and Reliance. Two research approaches have been proposed, the first of which employs the most basic method of directly applying stock closure data to the forecast model. The forecast method second method generates and employs ten technical features that reflect the stock price trend and range. The first method can provide us with an idea of the uptrend or downtrend, while the second method can provide us with actual stock behavior in the future. We observed significant improvements in prediction performance when compared to standard machine learning algorithms. The proposed method can predict stock movements in real time, making investments highly profitable and secure.

    Other economic elements which influence financial markets like government policies, interest rates, exchange rates, inflation, Twitter sentiment, or any hybrid approach as such can be used to create knowledge base or as an input to the prediction models. Another potential factor in determining the expected trend or price is stock volume.

    The focus of this study is on short-term forecasts. We forecasted future prices for the next 30 days. Other timeframes such as 5 days, 10 days and 15 days can be studied and compared. Long term forecasts could be chosen for further research or guidance and it may include quarterly stock performance reviews, sales,

profit returns, organizational stability, and other factors. Technical indicators based on short-term duration, such as SMA or EMA 10 days or 50 days, are available in this paper, but can also be researched and validated for longer timeframes if long-term predictions are desired.

Stock prices are known to be influenced by a variety of factors, including government policies, business operations, interest rates, and so on. Stock prices are affected by news of any of these variables. Natural disasters and other unpredictable events will have a significant impact on any excellent machine learning or deep learning model. As a result, a hybrid technique can be presented that takes all prospective aspects into account such as news, sentiments and technical indications, resulting in a more robust and accurate prediction system.